\documentstyle[prd,aps]{revtex}

\newcommand{\bea}{\begin{eqnarray}}
\newcommand{\eea}{\end{eqnarray}}

\begin{document}
%%%%%%%%%%%%%%%%%%%%%%%%%%%%%%%%%%%%%%%%%%%%%%%%%%%%%%%%%%%%%%%
\draft
%  For 2 column format.
%@twocolumnfalse\endcsname

%%%%%%%%%%%%%%%%%%%%%%%%%%%%%%%%%%%%%%%%%%%%%%%%%%%%%%%%%%%%%%%
\title{Why Newton's gravity is practically reliable \\
       in the large-scale cosmological simulations}

\author{Jai-chan Hwang${}^{(a)}$ and Hyerim Noh${}^{(b)}$}
\address{${}^{(a)}$ Department of Astronomy and Atmospheric Sciences,
                    Kyungpook National University, Taegu, Korea \\
         ${}^{(b)}$ Korea Astronomy and Space Science Institute,
                    Daejon, Korea \\
         E-mails: ${}^{(a)}$jchan@knu.ac.kr, ${}^{(b)}$hr@kasi.re.kr
         }
\date{\today}
\maketitle

%%%%%%%%%%%%%%%%%%%%%%%%%%%%%%%%%%%%%%%%%%%%%%%%%%%%%%%%%%%%%%%
\begin{abstract}

Until now, it has been common to use Newton's gravity to study the
non-linear clustering properties of the large-scale structures.
Without confirmation from Einstein's theory, however, it has been
unclear whether we can rely on the analysis, for example, near the
horizon scale. In this work we will provide a confirmation of using
Newton's gravity in cosmology based on relativistic analysis of
weakly non-linear situations to the third order in perturbations. We
will show that, except for the gravitational wave contribution, the
relativistic zero-pressure fluid equations perturbed to the second
order in a flat Friedmann background {\it coincide exactly} with the
Newtonian results. We will also present the {\it pure relativistic
correction} terms appearing in the third order. The third-order
correction terms show that these are the linear-order curvature
perturbation strength higher than the second-order
relativistic/Newtonian terms. Thus, the pure general relativistic
corrections in the third order are independent of the horizon scale
and are small in the large-scale due to the low-level temperature
anisotropy of the cosmic microwave background radiation. Since we
include the cosmological constant, our results are relevant to
currently favoured cosmology. As we prove that the Newtonian
hydrodynamic equations are valid in all cosmological scales to the
second order, and that the third-order correction terms are small,
our result has a practically important {\it implication} that one
can now use the large-scale Newtonian numerical simulation more
reliably as the simulation scale approaches and even goes beyond the
horizon.

\end{abstract}

%%%%%%%%%%%%%%%%%%%%%%%%%%%%%%%%%%%%%%%%%%%%%%%%%%%%%%%%%%%%%%%
%  For 2 column format.
%\vskip2pc]

%%%%%%%%%%%%%%%%%%%%%%%%%%%%%%%%%%%%%%%%%%%%%%%%%%%%%%%%%%%%%%%
%
% Introduction
%
%%%%%%%%%%%%%%%%%%%%%%%%%%%%%%%%%%%%%%%%%%%%%%%%%%%%%%%%%%%%%%%
\section{Introduction}

In order to interpret results from Einstein's gravity theory {\it
properly} we often need corresponding results in Newton's theory. On
the other hand, in order to use results from Newton's gravity theory
{\it reliably} we need confirmation from Einstein's theory. The
observed large-scale structures show non-linear processes are
working; see the 2dF galaxy redshift survey final data release
(Colless et al. 2003) and the SDSS data release 3 (Abazajian et al.
2004). Currently, studies of such structures are mainly based on
Newtonian physics in both analytical and numerical approaches (for
reviews, see Sahni \& Coles 1995; Bertschinger 1998; Bernardeau et
al. 2002; Cooray \& Sheth 2002). One may admit its incompleteness as
the simulation scale becomes bigger because, first, Newton's gravity
is an action-at-a-distance, i.e., the gravitational influence
propagates instantaneously thus violating causality. And, second,
Newton's theory is ignorant of the presence of horizon where the
relativistic effects are supposed to dominate: near horizon we have
$GM / (\lambda c^2) \sim {\lambda^2 / \lambda_H^2} \sim 1$ where
$\lambda_H \sim c/H$ is the dynamic horizon scale with $H$ Hubble's
constant. One other reason we may add is that Einstein's gravity
apparently has quite different structure from Newton's one. The
causality of gravitational interactions and consequent presence of
the horizon in cosmology are naturally taken into account in the
relativistic gravity theories where Einstein's gravity is the prime
example. In this work we will present the similarity and difference
between the two gravity theories in the weakly non-linear regimes in
cosmological situation.

In the literature, however, independently of such possible
shortcomings of Newton's gravity in the cosmological situation, the
sizes of Newtonian simulations, in fact, have already reached Hubble
horizon scale (Colberg et al. 2000; Jenkins et al. 2002; Evrard et
al. 2002; Bode \& Ostriker 2003; Dubinski et al. 2003; Park et al.
2005, in preparation). Common excuses often made by people working
in this active field of large-scale numerical simulation are, first
in the small scale one may rely on Newton's theory and, second, as
the scale becomes large the large-scale distribution of galaxies
looks linear in which case Einstein's gravity gives {\it the same}
result as the Newtonian one. In such a situation, in order to have
proper confirmation, we still need to investigate Einstein's case in
the non-linear or weakly non-linear situations. While the general
relativistic cosmological simulation is currently not available, in
this work, we will shed light on the situation by a perturbative
study of the non-linear regimes in Einstein's gravity. We will show
that even to the second order in perturbations, except for coupling
to gravitational waves, Einstein's gravity gives {\it the same}
equations known in Newton's theory and the pure relativistic
corrections appearing in the third-order perturbations are
independent of the horizon and are small. Thus, now our relativistic
analysis assures that Newton's gravity is practically reliable even
in the weakly non-linear regimes in cosmology. Such a comforting
conclusion comes from a thorough relativistic analysis of the weakly
non-linear situations to the third order in perturbations.  Despite
our simply expressed final conclusion the results still look
surprising and important. We set $c \equiv 1$.

%%%%%%%%%%%%%%%%%%%%%%%%%%%%%%%%%%%%%%%%%%%%%%%%%%%%%%%%%%%%%%%
%
% NL equations
%
%%%%%%%%%%%%%%%%%%%%%%%%%%%%%%%%%%%%%%%%%%%%%%%%%%%%%%%%%%%%%%%
\section{Fully non-linear equations and perturbations}

We start from the completely non-linear and covariant ($1+3$)
equations (Ehlers 1961; Ellis 1971, 1973). We need the energy
conservation equation and the Raychaudhury equation. {}For a
zero-pressure medium in the energy frame, we have (Noh \& Hwang
2004) \bea
   \tilde {\dot {\tilde \mu}} + \tilde \mu \tilde \theta
   &=& 0,
   \label{covariant-eq1} \\
   \tilde {\dot {\tilde \theta}} + {1 \over 3} \tilde \theta^2
       + \tilde \sigma^{ab} \tilde \sigma_{ab}
       - \tilde \omega^{ab} \tilde \omega_{ab}
       + 4 \pi G \tilde \mu - \Lambda
   &=& 0,
   \label{covariant-eq2}
\eea where $\Lambda$ is the cosmological constant; $\tilde \theta
\equiv \tilde u^a_{\;\; ;a}$ is the expansion scalar, and $\tilde
\sigma_{ab}$ and $\tilde \omega_{ab}$ are the shear and the rotation
tensors. Tildes indicate the covariant quantities based on the
spacetime metric $\tilde g_{ab}$. We have $\tilde {\dot {\tilde
\mu}} \equiv \tilde \mu_{,a} \tilde u^a$ and $\tilde {\dot {\tilde
\theta}} \equiv \tilde \theta_{,a} \tilde u^a$ which are the
covariant derivatives along $\tilde u^a$. {}From these we have \bea
   & & \left( {\tilde {\dot {\tilde \mu}} \over \tilde \mu}
       \right)^{\tilde \cdot}
       - {1 \over 3}
       \left( {\tilde {\dot {\tilde \mu}} \over \tilde \mu} \right)^2
       - \tilde \sigma^{ab} \tilde \sigma_{ab}
       + \tilde \omega^{ab} \tilde \omega_{ab}
       - 4 \pi G \tilde \mu
       + \Lambda
       = 0.
   \label{covariant-eq3}
\eea Equations (\ref{covariant-eq1})-(\ref{covariant-eq3}) are valid
to all orders, i.e., these equations are {\it fully non-linear} and
still covariant. Equations
(\ref{covariant-eq1})-(\ref{covariant-eq3}) are not complete: to the
second and higher order perturbations we will need other equations
in Einstein's theory.

We consider the {\it scalar-} and {\it tensor-type} perturbations in
the Friedmann background without pressure; we ignore the vector-type
perturbation (rotation) because it always decays in the expanding
phase even to the second order (Noh \& Hwang 2004). As the metric we
take \bea
   ds^2
   &=& - \left( 1 + 2 \alpha \right) d t^2
       - 2 a \beta_{,\alpha} d t d x^\alpha
       + a^2 \left[ g^{(3)}_{\alpha\beta} \left( 1 + 2 \varphi \right)
       + 2 \gamma_{,\alpha|\beta}
       + 2 C^{(t)}_{\alpha\beta} \right] d x^\alpha d x^\beta,
   \label{metric}
\eea where $a(t)$ is the scale factor; $\alpha$, $\beta$, $\gamma$
and $\varphi$ are spacetime dependent scalar-type perturbed-order
variables; $C^{(t)}_{\alpha\beta}$ is the transverse and tracefree
tensor-type metric perturbation (gravitational waves). We take the
metric convention in Bardeen (1988) extended to the third order (Noh
\& Hwang 2004).  The Greek and Latin indices indicate the space and
spacetime indices, respectively; the spatial indices of perturbed
order variables are raised and lowered by $g^{(3)}_{\alpha\beta}$
which becomes $\delta_{\alpha\beta}$ if we take Cartesian
coordinates in the flat Friedmann background. A vertical bar
indicates a covariant derivative based on $g^{(3)}_{\alpha\beta}$.
We will take $\gamma \equiv 0$ as the spatial gauge condition which
makes all the remaining variables spatially gauge-invariant to the
all orders in perturbations (Noh \& Hwang 2004).

The fluid quantities are ordinarily defined based on the fluid
four-vector $\tilde u_a$ in the energy-frame. Our comoving gauge
condition takes vanishing flux $\tilde q_a \equiv 0$ (the
energy-frame), and $\tilde u_\alpha \equiv 0$ for the fluid
four-vector; here, we ignored the vector-type perturbation. Thus,
the fluid four-vector coincides with the normal-frame four-vector
$\tilde n_a$ which has $\tilde n_\alpha \equiv 0$. The condition
$\tilde u_\alpha = 0$ implies vanishing rotation tensor $\tilde
\omega_{ab} = 0$. We lose no generality by imposing the gauge
condition. Since the comoving gauge condition fixes the temporal
gauge mode completely the remaining variables under this gauge
condition are equivalently gauge-invariant; this is the case to all
orders in perturbations (Noh \& Hwang 2004). In our gauge condition
the energy-momentum tensor becomes \bea
   & & \tilde T^0_0 = - \tilde \mu, \quad
       \tilde T^0_\alpha = 0 = \tilde T^\alpha_\beta,
   \label{Tab}
\eea where $\tilde \mu$ is the energy density.

%%%%%%%%%%%%%%%%%%%%%%%%%%%%%%%%%%%%%%%%%%%%%%%%%%%%%%%%%%%%%%%
%
% BG order
%
%%%%%%%%%%%%%%%%%%%%%%%%%%%%%%%%%%%%%%%%%%%%%%%%%%%%%%%%%%%%%%%
\section{Background and linear perturbations}

To the background order, we have $\tilde \mu = \mu$ and $\tilde
\theta = 3 {\dot a \over a}$ where an overdot indicates a time
derivative based on $t$. Equations (\ref{covariant-eq1}) and
(\ref{covariant-eq2}) give \bea
   \dot \mu + 3 {\dot a \over a} \mu
   &=& 0,
   \\
   3 {\ddot a \over a} + 4 \pi G \mu - \Lambda
   &=& 0.
\eea Combining these equations we have the Friedmann equation
 \bea
   & & {\dot a^2 \over a^2} = {8 \pi G \over 3} \mu
       - {{\rm const.} \over a^2} + {\Lambda \over 3},
   \label{BG}
\eea with $\mu \propto a^{-3}$. This equation was first derived
based on Einstein's gravity by Friedmann (1922, 1924) and Robertson
(1929), and Newtonian study followed later by Milne (1934) and
McCrea \& Milne (1934). In the Newtonian context the relativistic
energy density $\mu$ can be identified with the mass density
$\varrho$. The ``{\rm const.}'' part is interpreted as the spatial
curvature ($K$) in Einstein's gravity (Friedmann 1922, 1924) and the
total energy in Newton's gravity (McCrea \& Milne 1934).

%%%%%%%%%%%%%%%%%%%%%%%%%%%%%%%%%%%%%%%%%%%%%%%%%%%%%%%%%%%%%%%
%
% Linear order
%
%%%%%%%%%%%%%%%%%%%%%%%%%%%%%%%%%%%%%%%%%%%%%%%%%%%%%%%%%%%%%%%
To the linear-order perturbations in the metric and energy-momentum
variables, we introduce \bea
   & & \tilde \mu \equiv \mu + \delta \mu, \quad
       \tilde \theta \equiv 3 {\dot a \over a} + \delta \theta,
   \label{perturbations}
\eea where $\mu$ and $\delta \mu$ are the background and perturbed
energy density, respectively, and $\delta \theta$ is the perturbed
part of expansion scalar. We emphasize that our spatial $\gamma = 0$
gauge and temporal comoving gauge conditions defined above eq.\
(\ref{Tab}) fix the spatial and temporal gauge degrees of freedom
completely. Thus, all variables in these gauge conditions are
equivalently gauge invariant to the linear order: i.e., each of the
variables has a unique corresponding gauge-invariant combination
(Bardeen 1988; Hwang 1991). It is important to point out that the
above two statements are valid even in the second and all higher
order perturbations, see \S VI in Noh \& Hwang (2004). To the
background order we already identified $\mu \equiv \varrho$. Now, to
the linear order we {\it identify} \bea
   & & \delta \mu \equiv \delta \varrho, \quad
       \delta \theta \equiv {1 \over a} \nabla \cdot {\bf u}.
   \label{identification}
\eea To the linear order the perturbed parts of eqs.\
(\ref{covariant-eq1}) and (\ref{covariant-eq2}) give \bea
   \dot \delta + {1 \over a} \nabla \cdot {\bf u}
   &=& 0,
   \label{linear-eq1} \\
   {1 \over a} \nabla \cdot \left( \dot {\bf u}
       + {\dot a \over a} {\bf u} \right) + 4 \pi G \mu \delta
   &=& 0.
   \label{linear-eq2}
\eea Combining these equations we have \bea
   & & \ddot \delta + 2 {\dot a \over a} \dot \delta - 4 \pi G \mu \delta
       = 0,
   \label{pert}
\eea which is the well known density perturbation equation in both
relativistic and Newtonian contexts; we set $\delta \equiv {\delta
\mu / \mu}$. This equation was first derived based on Einstein's
gravity by Lifshitz (1946), and Newtonian study followed later by
Bonnor (1957). Notice that the relativistic result is {\it
identical} to the Newtonian result.  The gravitational wave
perturbation present in the relativistic theory simply decouples
from the density perturbation and follows the wave equation
(Lifshitz 1946) \bea
   & & \ddot C^{(t)}_{\alpha\beta}
       + 3 {\dot a \over a} \dot C^{(t)}_{\alpha\beta}
       - {\Delta - 2 K \over a^2} C^{(t)}_{\alpha\beta} = 0,
   \label{GW-eq}
\eea where $K$ is the sign of the background spatial curvature.

It is curious to notice that in both the expanding world model and
its linear structures the first studies were made in the context of
Einstein's gravity (Friedmann 1922; Lifshitz 1946), and the much
simpler and, in hindsight, more intuitive Newtonian studies followed
later (Milne 1934; Bonnor 1957). Perhaps these historical
developments reflect that people did not have confidence in using
Newton's gravity in cosmology before the result was already known
in, and the method was ushered by, Einstein's gravity.  This is also
reflected in the historical development of modern cosmology which
began only after the advent of Einstein's gravity theory (Einstein
1917).
%\footnote{
%          The only preceding cosmologically relevant discussions can be found
%          in Newton's correspondences to Bentley in 1692 (Newton 1692):
%          ``{\sl But if the matter was evenly disposed throughout an
%                 infinite space, it could never convene into one mass;
%                 but some of it would convene into one mass and some into
%                 another, so as to make an infinite number of great masses,
%                 scattered at great distances from one to another throughout
%                 all that infinite space.}''
%}.
{}Furthermore, it is known in the literature that the results in
Newtonian cosmology are, in fact, guided ones by previously known
relativistic results; i.e., without the guidance of the relativistic
analyses Newtonian theory could have lead to other results (Layzer
1954; Lemons 1988). It may be also true that only after having a
Newtonian counterpart we could understand better what the often
arcane relativistic analysis shows. {}For the second-order
perturbations, however, the history is different from the two
previous cases. Currently we only have the Newtonian result known in
the literature. Thus, the result only known in Newton's gravity
still awaits confirmation from Einstein's theory.  Here, we are
going to fill the gap by presenting the much needed relativistic
confirmation to the second order and the pure relativistic
corrections start appearing from the third order.

Although eq.\ (\ref{pert}) is also valid with general spatial
curvature, in the following we consider the {\it flat} background
only. As we include the cosmological constant $\Lambda$, however,
our zero-pressure background and perturbations describe remarkably
well the current expanding stage of our universe and its large-scale
structures (Spergel et al. 2003; Tegmark, et al. 2004), which are
believed to be in the near linear stage. In the small scale,
however, the structures are apparently in non-linear stage, and even
in the large scale weakly non-linear study is needed. Until now,
such a weakly non-linear stage has been studied in Newton's gravity
only.  In the following we plan to investigate whether such a usage
of Newtonian gravity in handling the large-scale structure can be
justified in the relativistic standpoint by studying the
relativistic behaviours of higher-order perturbations.

%%%%%%%%%%%%%%%%%%%%%%%%%%%%%%%%%%%%%%%%%%%%%%%%%%%%%%%%%%%%%%%
%
% To second order
%
%%%%%%%%%%%%%%%%%%%%%%%%%%%%%%%%%%%%%%%%%%%%%%%%%%%%%%%%%%%%%%%
\section{Second-order perturbations and Newtonian correspondence}

Now, we consider equations perturbed to the second order in the
metric and the energy-momentum tensor. Even to the second order we
introduce perturbations as in eq.\ (\ref{perturbations}) which are
always allowed. We will also {\it take} the same identifications
made in eq.\ (\ref{identification}); this point will be justified by
our results soon. To the second order the perturbed parts of eqs.\
(\ref{covariant-eq1}) and (\ref{covariant-eq2}) give (Hwang \& Noh
2005a) \bea
   \dot \delta + {1 \over a} \nabla \cdot {\bf u}
   &=& - {1 \over a} \nabla \cdot \left( \delta {\bf u} \right),
   \label{dot-delta-eq-2nd} \\
   {1 \over a} \nabla \cdot \left( \dot {\bf u}
       + {\dot a \over a} {\bf u} \right)
       + 4 \pi G \mu \delta
   &=& - {1 \over a^2} \nabla \cdot
       \left( {\bf u} \cdot \nabla {\bf u} \right)
       - \dot C^{(t)\alpha\beta} \left( {2 \over a} \nabla_\alpha u_\beta
       + \dot C^{(t)}_{\alpha\beta} \right),
   \label{dot-delta-u-eq-2nd}
\eea where the gravitational wave part comes from the shear term in
eq.\ (\ref{covariant-eq2}) (Noh \& Hwang 2004, 2005; Hwang \& Noh
2005a) and it follows eq.\ (\ref{GW-eq}); in order to derive these
equations we also used the $\tilde G^0_\alpha$-component (momentum
constraint) of Einstein's field equations. By combining these
equations we have \bea
   \ddot \delta + 2 {\dot a \over a} \dot \delta - 4 \pi G \mu \delta
   &=& - {1 \over a^2} {\partial \over \partial t}
       \left[ a \nabla \cdot \left( \delta {\bf u} \right) \right]
       + {1 \over a^2} \nabla \cdot \left( {\bf u} \cdot
       \nabla {\bf u} \right)
       + \dot C^{(t)\alpha\beta} \left( {2 \over a} \nabla_\alpha u_\beta
       + \dot C^{(t)}_{\alpha\beta} \right),
   \label{ddot-delta-eq-2nd}
\eea which also follows from eq.\ (\ref{covariant-eq3}). Equations
(\ref{dot-delta-eq-2nd})-(\ref{ddot-delta-eq-2nd}) are our extension
of eqs.\ (\ref{linear-eq1})-(\ref{pert}) to the second-order
perturbations in Einstein's theory. We will show that, except for
gravitational waves, {\it exactly the same} equations also follow
from Newton's theory. The presence of the gravitational waves,
however, can be regarded as one of the truly relativistic effects of
gravitation.

%%%%%%%%%%%%%%%%%%%%%%%%%%%%%%%%%%%%%%%%%%%%%%%%%%%%%%%%%%%%%%%
%
% Newtonian fully non-linear perturbations
%
%%%%%%%%%%%%%%%%%%%%%%%%%%%%%%%%%%%%%%%%%%%%%%%%%%%%%%%%%%%%%%%
In the Newtonian context, the mass and the momentum conservations,
and Poisson's equation give (Peebles 1980) \bea
   \dot \delta + {1 \over a} \nabla \cdot {\bf u}
   &=& - {1 \over a} \nabla \cdot \left( \delta {\bf u} \right),
   \label{mass-conservation} \\
   \dot {\bf u} + {\dot a \over a} {\bf u}
   + {1 \over a} \nabla \delta \Phi
   &=& - {1 \over a} {\bf u} \cdot \nabla {\bf u},
   \label{momentum-conservation} \\
   {1\over a^2} \nabla^2 \delta \Phi
   &=& 4 \pi G \varrho \delta,
   \label{Poisson-eq}
\eea where $\delta \Phi$ is the perturbed gravitational potential,
${\bf u}$ is the perturbed velocity, and $\delta \equiv {\delta
\varrho / \varrho}$. Equation (\ref{dot-delta-eq-2nd}) is the same
as eq.\ (\ref{mass-conservation}), and eq.\
(\ref{dot-delta-u-eq-2nd}) ignoring gravitational waves follows from
eqs.\ (\ref{momentum-conservation}) and (\ref{Poisson-eq}). Thus,
eq.\ (\ref{ddot-delta-eq-2nd}) also naturally follows in Newton's
theory (Peebles 1980). This shows the {\it exact}
relativistic-Newtonian correspondence to the second order, except
for the gravitational wave contribution which is a pure relativistic
effect. This also justifies our identifications made in eq.\
(\ref{identification}) to the second order. Although we identified
the relativistic density and velocity perturbation variables we {\it
cannot} identify a relativistic variable which corresponds to
$\delta \Phi$ to the second order (Hwang \& Noh 2005a).  We believe
this can be understood naturally because Poisson's equation indeed
reveals the action-at-a-distance nature and the static nature of
Newton's gravity theory compared with Einstein's gravity (Fock 1964;
Rindler 1977). Poisson's equation was formulated in 1812 which was
125 years after the publication of Newton's Principia in 1687.
Notice that eqs.\ (\ref{mass-conservation})-(\ref{Poisson-eq}) are
valid to {\it fully non-linear} order. In our relativistic case,
however, eqs.\ (\ref{dot-delta-eq-2nd})-(\ref{ddot-delta-eq-2nd})
are valid only to the second order in perturbations.

%%%%%%%%%%%%%%%%%%%%%%%%%%%%%%%%%%%%%%%%%%%%%%%%%%%%%%%%%%%%%%%
%
% Third order perturbations
%
%%%%%%%%%%%%%%%%%%%%%%%%%%%%%%%%%%%%%%%%%%%%%%%%%%%%%%%%%%%%%%%
\section{Third-order perturbations and pure relativistic
corrections}

Since the zero-pressure Newtonian system is exact to the second
order in non-linearity, all non-vanishing third and higher order
perturbation terms in the relativistic analysis can be regarded as
the pure relativistic corrections. Thus, we have a clear reason to
go to the third order which was not previously attempted. {}For
simplicity we ignore gravitational wave contribution; for a complete
presentation, see Hwang \& Noh (2005b). Based on our success in the
second-order perturbations, we continue {\it identifying} eq.\
(\ref{identification}) is valid even to the {\it third order} and
will {\it take} the consequent additional third-order terms as the
pure relativistic corrections. To the third order the perturbed
parts of eqs.\ (\ref{covariant-eq1}) and (\ref{covariant-eq2}) give
(Hwang \& Noh 2005b): \bea
   \dot \delta + {1 \over a} \nabla \cdot {\bf u}
   &=& - {1 \over a} \nabla \cdot \left( \delta {\bf u} \right)
   \nonumber \\
   & &
       + {1 \over a} \left[ 2 \varphi {\bf u}
       - \nabla \left( \Delta^{-1} X \right) \right] \cdot \nabla \delta,
   \label{delta-eq-3rd} \\
   {1 \over a} \nabla \cdot \left( \dot {\bf u}
       + {\dot a \over a} {\bf u} \right)
       + 4 \pi G \mu \delta
   &=& - {1 \over a^2} \nabla \cdot \left( {\bf u}
       \cdot \nabla {\bf u} \right)
   \nonumber \\
   & &
       - {2 \over 3 a^2} \varphi
       {\bf u} \cdot \nabla \left( \nabla \cdot {\bf u} \right)
       + {4 \over a^2} \nabla \cdot \left[ \varphi
       \left( {\bf u} \cdot \nabla {\bf u}
       - {1 \over 3} {\bf u} \nabla \cdot {\bf u} \right) \right]
   \nonumber \\
   & &
       - {\Delta \over a^2}
       \left[ {\bf u} \cdot \nabla \left( \Delta^{-1} X \right) \right]
       + {1 \over a^2} {\bf u} \cdot \nabla X
       + {2 \over 3a^2} X \nabla \cdot {\bf u},
   \label{u-eq-3rd}
\eea where \bea
   X
   &\equiv&
       2 \varphi \nabla \cdot {\bf u}
       - {\bf u} \cdot \nabla \varphi
       + {3 \over 2} \Delta^{-1} \nabla \cdot
       \left[ {\bf u} \cdot \nabla \left( \nabla \varphi \right)
       + {\bf u} \Delta \varphi \right].
   \label{X-eq-3rd}
\eea In order to derive these equations we also used the $\tilde
G^0_\alpha$-component of Einstein's field equations.  Equations
(\ref{delta-eq-3rd}) and (\ref{u-eq-3rd}) extend eqs.\
(\ref{dot-delta-eq-2nd}) and (\ref{dot-delta-u-eq-2nd}) to the third
order. By combining eqs.\ (\ref{delta-eq-3rd}) and (\ref{u-eq-3rd})
we can derive \bea
   \ddot \delta + 2 {\dot a \over a} \dot \delta
       - 4 \pi G \mu \delta
   &=& - {1 \over a^2} {\partial \over \partial t}
       \left[ a \nabla \cdot \left( \delta {\bf u} \right) \right]
       + {1 \over a^2} \nabla \cdot \left( {\bf u}
       \cdot \nabla {\bf u} \right)
   \nonumber \\
   & &
       + {1 \over a^2} {\partial \over \partial t}
       \left\{ a \left[ 2 \varphi {\bf u}
       - \nabla \left( \Delta^{-1} X \right) \right] \cdot \nabla \delta
       \right\}
       + {2 \over 3 a^2} \varphi
       {\bf u} \cdot \nabla \left( \nabla \cdot {\bf u} \right)
       - {4 \over a^2} \nabla \cdot \left[ \varphi
       \left( {\bf u} \cdot \nabla {\bf u}
       - {1 \over 3} {\bf u} \nabla \cdot {\bf u} \right) \right]
   \nonumber \\
   & &
       + {\Delta \over a^2}
       \left[ {\bf u} \cdot \nabla \left( \Delta^{-1} X \right) \right]
       - {1 \over a^2} {\bf u} \cdot \nabla X
       - {2 \over 3a^2} X \nabla \cdot {\bf u},
   \label{density-eq-3rd}
\eea which extends eq.\ (\ref{ddot-delta-eq-2nd}) to the third
order. The last two lines of eq.\ (\ref{density-eq-3rd}) are pure
third-order terms. The variable $\varphi$ is a perturbed-order
metric variable in eq.\ (\ref{metric}) in our comoving gauge
condition.

The third-order correction terms in eqs.\
(\ref{delta-eq-3rd})-(\ref{density-eq-3rd}) reveal that all of them
are simply of $\varphi$-order higher than the second-order terms.
Thus, the pure general relativistic effects are at least
$\varphi$-order higher than the relativistic/Newtonian ones in the
second order. Our $\varphi$ is related to the perturbed three-space
curvature (in our comoving gauge) and dimensionless (Bardeen 1980).
As we mentioned earlier, $\varphi$ in the comoving gauge is the same
as a unique gauge-invariant combination. To the linear order such a
combination was first introduced by Field \& Shepley (1968). {}For
an explicit form of the combination to the second order, see eq.\
(281) in Noh \& Hwang (2004). Notice that we only need the behavior
of $\varphi$ to the linear order. To the linear order, in terms of
known Newtonian variables we have (Hwang \& Noh 2005b) \bea
   & & \varphi = - \delta \Phi
       + \dot a \Delta^{-1} \nabla \cdot {\bf u},
\eea and it satisfies (Hwang \& Noh 1999a) \bea
   & & \dot \varphi = 0,
\eea thus $\varphi = C ({\bf x})$ with {\it no} decaying mode; this
is true considering the presence of the cosmological constant. {}For
$\Lambda = 0$, the temperature anisotropy of cosmic microwave
background radiation gives (Sachs \& Wolfe 1967; Hwang \& Noh 1999b)
\bea
   & & {\delta T \over T} \sim {1 \over 3} \delta \Phi
       \sim {1 \over 5} \varphi.
   \label{SW}
\eea  The observations of cosmic microwave background radiation
give ${\delta T / T} \sim 10^{-5}$ (Smoot et al. 1992; Spergel, et
al. 2003), thus \bea
   & & \varphi \sim 5 \times 10^{-5},
   \label{CMB-constraint}
\eea in the large-scale limit near the horizon scale where
$GM/(\lambda c^2) \sim \lambda^2/\lambda_H^2$ approaches unity.
Therefore, to the third order, the pure relativistic corrections are
{\it independent} of the horizon scale and depend on the
linear-order curvature $\varphi$ ($\sim$ gravitational potential
$\delta \Phi$) perturbation strength {\it only}, and are small. That
is, compared with the second-order terms, the third-order correction
terms in eqs.\ (\ref{delta-eq-3rd})-(\ref{X-eq-3rd}) only involve
$\varphi$, and do not contain terms like $(aH)^{-1} \nabla \varphi$,
etc.

%%%%%%%%%%%%%%%%%%%%%%%%%%%%%%%%%%%%%%%%%%%%%%%%%%%%%%%%%%%%%%%
%
% Discussion
%
%%%%%%%%%%%%%%%%%%%%%%%%%%%%%%%%%%%%%%%%%%%%%%%%%%%%%%%%%%%%%%%
\section{Discussion}

We have shown that to the second order, except for the gravitational
wave contribution, the zero-pressure relativistic cosmological
perturbation equations can be exactly identified with the known
equations in Newton's theory. As a consequence, to the second order,
we identified correct relativistic variables which can be
interpreted as density $\delta \mu$ and velocity $\delta \theta$
perturbations in eq.\ (\ref{identification}), and we showed that to
the second order the Newtonian hydrodynamic equations remain valid
in all cosmological scales including the super-horizon scale.  It
might as well happen that our relativistic results give relativistic
correction terms appearing to the second order which become
important as we approach and go beyond the horizon scale which are
strongly relativistic regimes. Our results show that there are no
such correction terms appearing to the second order, and except for
gravitational waves, the correspondence is exact to that order.
Ignoring gravitational waves, the pure relativistic correction
terms, however, start appearing from the third order. Our study
shows that to the third order the correction terms only involve
$\varphi$ which is again independent of the the horizon scale and is
small in the large scale.

In the non-linear clustered regions we may have $\varphi \sim \delta
\Phi \sim GM / (R c^2)$ where $M$ and $R$ are characteristic mass
and length scales involved. In such clustered regimes the
post-Newtonian approximation would complement our non-linear
perturbation approach presented here. The post-Newtonian
approximation takes $v/c$-expansion when the motions are slow $v/c
\ll 1$, and gravity is weak $GM/(Rc^2) \ll 1$. A complementary
result, showing the relativistic-Newtonian correspondence in the
Newtonian limit of the post-Newtonian approach, can be found in
Kofman \& Pogosyan (1995), (see also Bertschinger \& Hamilton 1994;
Ellis \& Dunsby 1997). In fact, the Newtonian hydrodynamic equations
naturally appear in the zeroth-order post-Newtonian approximation
(Chandrasekhar 1965). Recently, we presented the fully nonlinear
cosmological hydrodynamic equations with first-order post-Newtonian
correction terms (Hwang, Noh \& Puetzfeld 2005); we showed that
these correction terms have typically $GM/(Rc^2) \sim v^2/c^2 \sim
10^{-5}$ order smaller than the Newtonian terms in the non-linearly
clustered regions.

Therefore, our general relativistic results allow us to draw the
following important practical conclusion which is stated in our
title. As we prove that the Newtonian hydrodynamic equations are
valid on all cosmological scales to the second order, and that the
third-order pure relativistic correction terms are small and
independent of the horizon, one can now use the large-scale
Newtonian numerical simulation more reliably as the simulation scale
approaches and even goes beyond the horizon. The fluctuations near
the horizon scale are supposed to be linear or weakly nonlinear;
otherwise, it is difficult to introduce the spatially homogeneous
and isotropic background world model which is the basic assumption
of the modern cosmology. In the small-scale but fully non-linear
stage, the post-Newtonian approximation also shows that the
relativistic correction terms are small, thus the Newtonian
simulations can be trusted again. The sub-horizon scale Newtonian
non-linear inhomogeneities are not supposed to affect the
homogeneous and isotropic background world model (Siegel \& Fry
2005). The other side of this conclusion is that it might be
difficult to find testable signatures of Einstein's gravity theory
based on such large-scale weakly non-linear structures (with
relativistic corrections) or small-scale fully non-linear structures
(with post-Newtonian corrections). However, it would be interesting
to find cosmological situations where the pure relativistic
correction terms in eqs.\
(\ref{delta-eq-3rd})-(\ref{density-eq-3rd}) or the first-order
post-Newtonian corrections terms derived in Hwang, Noh \& Puetzfeld
(2005) could have observationally distinguishable consequences.
Since our equations include the cosmological constant our equations
and conclusions are relevant to the currently favoured world models.

In our relativistic-Newtonian correspondence to the second order,
the relativistic equations are identified with the continuity
equation and the divergence of the Euler equation replacing the
Newtonian gravitational potential using Poisson's equation. It is
important to be remember that we showed the relativistic-Newtonian
correspondence for the density and velocity perturbations, but not
for the gravitational potential. Therefore, although our result
assures that one can trust cold dark matter simulations  at {\it
all} scales for the density and velocity fields, it does {\it not}
imply that one can trust the Newtonian simulations for effects
involving the gravitational potential, like the weak gravitational
lensing effects. In order to handle the lensing effects properly we
often require an extra factor of two which, indeed, comes from the
post-Newtonian effects.

Our relativistic-Newtonian correspondence to the second order
perturbation is valid for the scalar-type perturbation {\it
assuming} a single component zero-pressure irrotational fluid in the
flat cosmological background. Dropping any of these conditions could
potentially lead to relativistic corrections. The genuine
relativistic correction terms appear as we consider the
gravitational waves to the second order. We showed that pure
relativistic correction terms in the scalar-type perturbation appear
in the third order; we showed that these correction terms do not
involve the horizon scale and are small in our observable patch of
the universe. Extensions to include the pressure, the rotation, the
non-flat background, and the multi-component situation will be
investigated in future occasions.

%%%%%%%%%%%%%%%%%%%%%%%%%%%%%%%%%%%%%%%%%%%%%%%%%%%%%%%%%%%%%%%
%
% Acknowledgments
%
%%%%%%%%%%%%%%%%%%%%%%%%%%%%%%%%%%%%%%%%%%%%%%%%%%%%%%%%%%%%%%%
\subsection*{Acknowledgments}

HN was supported by grants No. R04-2003-10004-0 from the Basic
Research Program of the Korea Science and Engineering Foundation. JH
was supported by the Korea Research Foundation Grant No.
2003-015-C00253.

%%%%%%%%%%%%%%%%%%%%%%%%%%%%%%%%%%%%%%%%%%%%%%%%%%%%%%%%%%%%%%%
%
% References
%
%%%%%%%%%%%%%%%%%%%%%%%%%%%%%%%%%%%%%%%%%%%%%%%%%%%%%%%%%%%%%%%

%%%%%%%%%%%%%%%%%%%%%%%%%%%%%%%%%%%%%%%%%%%%%%%%%%%%%%%%%%%%%%%

\begin{thebibliography}{99}
\bibitem{2dF}
         Abazajian K., et al. 2004, in http://www.sdss.org/dr3/
\bibitem{Bardeen-1980}
         Bardeen J. M., 1980, Phys. Rev. D, 22, 1882
\bibitem{Bardeen-1988}
         Bardeen J. M., 1988,
                  in Fang, L., Zee, A., eds. Particle Physics and Cosmology,
                  Gordon and Breach, London, p. 1
\bibitem{Bernardeau}
         Bernardeau F., Colombi S., Gaztanaga E.,
                     Scoccimarro, R., 2002 Phys. Rep., 367, 1
\bibitem{Bertschinger2}
         Bertschinger E., 1998, ARA\&A, 36, 599
\bibitem{Bertschinger}
         Bertschinger E., Hamilton A. J. S., 1994, ApJ, 435, 1
\bibitem{Bode}
         Bode, P, Ostriker, J. P., 2003, ApJS, 145, 1
\bibitem{Bonnor-1957}
         Bonnor W. B., 1957, MNRAS, 117, 104
\bibitem{Chandrasekhar-1965}
         Chandrasekhar S., 1965, Astrophys. J., 142, 1488
\bibitem{simulation}
         Colberg J. M., et al., 2000, MNRAS, 319, 209
\bibitem{Colless-etal-2003}
         Colless M., et al., 2003, preprint (astro-ph/0306581)
\bibitem{Cooray}
         Cooray A., Sheth, R., 2002, Phys. Rep., 372, 1
\bibitem{Dubinski}
         Dubinski J., Kim J., Park C., Humble R.,
                   2003, New Astron., 9, 111
\bibitem{covariant}
         Ehlers J., 1961, Proceedings of the mathematical-natural science of
                 the Mainz academy of science and literature, Nr. 11, 792;
                 translated in 1993, Gen. Rel. Grav., 25, 1225
\bibitem{Einstein}
         Einstein A., 1917, Sitzungsberichte der Preussischen Akad. d. Wiss.,
                   1917, 142;
                   translated in Bernstein J., Feinberg G., eds, 1986,
                   Cosmological-constants: papers in modern cosmology,
                   Columbia Univ. Press, New York, p. 16
\bibitem{Ellis-1971}
         Ellis G. F. R., 1971,
                in Sachs R. K., ed,
                General relativity and cosmology, Proceedings
                of the international summer school of physics
                Enrico Fermi course 47, Academic Press, New York, p. 104
\bibitem{Ellis-1973}
         Ellis G. F. R., 1973,
                in E. Schatzmann eds, Cargese Lectures in Physics,
                Gorden and Breach, New York, p. 1
\bibitem{Ellis-Dunsby-1997}
         Ellis G. F. R., Dunsby P. K. S., 1997, ApJ, 479, 97
\bibitem{Evrard}
         Evrard A. E., et al., 2002, ApJ, 573, 7
\bibitem{FS}
         Field G. B., Shepley, L. C., 1968, Ap\&SS, 1, 309
\bibitem{Fock-1964}
         Fock V., 1964, The theory of space, time, and gravitation,
                  2nd revised edition, The MacMillan Company, New York
\bibitem{Friedmann-1922}
         Friedmann A. A., 1922, Zeitschrift f\"ur Physik, 10, 377;
                    translated in Bernstein J., Feinberg G., eds,
                    1986, Cosmological-constants: papers in modern cosmology,
                    Columbia Univ. Press, New York, p. 49
\bibitem{Friedmann-1924}
         Friedmann A. A., 1924, Zeitschrift f\"ur Physik, 21, 326;
                    translated in Bernstein J., Feinberg G., eds.
                    1986, Cosmological-constants: papers in modern cosmology,
                    Columbia Univ. Press, New York, p. 59
\bibitem{H-1991}
         Hwang J., 1991 ApJ, 375, 443
\bibitem{HN-Newtonian-1999}
         Hwang J., Noh H., 1999a, Gen. Rel. Grav., 31, 1131
\bibitem{HN-SW1}
         Hwang J., Noh H., 1999b, Phys. Rev. D, 59, 067302
\bibitem{HN-second-order}
         Hwang J., Noh H., 2005a, Phys. Rev. D, in press,
                   preprint (gr-qc/0412128)
\bibitem{HN-third-order}
         Hwang J., Noh H., 2005b, Phys. Rev. D, in press,
                   preprint (gr-qc/0412129)
\bibitem{HNP}
         Hwang J., Noh H., Puetzfeld D., 2005, Phys. Rev. D submitted,
                   preprint (astro-ph/0507085)
\bibitem{Jenkins}
         Jenkins A., et al., 2001, MNRAS, 321, 372
\bibitem{Kofman-Pogosyan-1995}
         Kofman L., Pogosyan D., 1995, ApJ, 442, 30
\bibitem{Layzer-1954}
         Layzer D., 1954, AJ, 59, 268
\bibitem{Lemons-1988}
         Lemons D. S., 1988, Am. J. Phys., 56, 502
\bibitem{Lifshitz-1946}
         Lifshitz E. M., 1946, J. Phys. (USSR), 10, 116
\bibitem{McCrea}
         McCrea W. H., Milne E. A., 1934, Quart. J. Math., 5, 73
\bibitem{Milne-1934}
         Milne E. A., 1934, Quart. J. Math., 5, 64
%\bibitem{Newton}
%         Newton I., 1692, Four Letters to Richard Bentley,
%                 reprinted in Munitz M. K., ed. 1957,
%                 Theories of the universe, The Free Press, London, p. 211
\bibitem{NL}
         Noh H., Hwang J., 2004, Phys. Rev. D, 69, 104011
\bibitem{Second-CQG}
         Noh H., Hwang J., 2005, Class. Quant. Grav., in press,
                 preprint (gr-qc/0412127)
%\bibitem{Park}
%         Park C., Kim J., Gott J. R., 2005, in preparation
\bibitem{Peebles-1980}
         Peebles P. J. E., 1980, The large-scale structure of the universe,
                  Princeton Univ. Press, Princeton
\bibitem{Rindler}
         Rindler, W., 1980, Essential relativity, Springer-Verlag, New York
\bibitem{Robertson}
         Robertson H. P., 1929, Proceedings of the National Academy of
                    Science, 15, 822
\bibitem{SW-1967}
         Sachs R. K., Wolfe A. M., 1967 ApJ, 147, 73
\bibitem{Sahni}
         Sahni V., Coles P., 1995, Phys. Rep., 262, 1
\bibitem{Siegel-Fry-2005}
         Siegel, E. R., Fry, J. N., 2005, preprint (astro-ph/0504421)
\bibitem{CMB}
         Smoot G. F., et al., 1992, ApJ, 396, L1
\bibitem{Spergel}
         Spergel D. N., et al., 2003, ApJS, 148, 175
\bibitem{Tegmark}
         Tegmark M., et al., 2004, Phys. Rev. D, 69, 103501
\end{thebibliography}
\end{document}